\documentclass[superscriptaddress,twocolumn,showpacs,preprintnumbers,amsmath,amssymb]{revtex4}
\newcommand{\be}{\begin{equation}}
\newcommand{\bea}{\begin{eqnarray}}
\newcommand{\eea}{\end{eqnarray}}
\newcommand{\ba}{\begin{array}}
\newcommand{\ea}{\end{array}}
\newcommand{\ee}{\end{equation}}

\begin{document}
\title{The Hubble Web:\\
 The Dark Matter Problem and Cosmic Strings}
\author{Stephon Alexander}

\affiliation{Departments of Physics and Astronomy\\
Institute for Gravity and Geometry,\\
The Pennsylvania State University,\\
University Park, PA 16802 \\ }
\email{sha3@psu.edu}

\date{\today}

\begin{abstract}
{I propose a reinterpretation of cosmic dark matter in which a rigid network of cosmic strings formed at the end of inflation.  The cosmic strings fulfill three functions:  At recombination they provide an accretion mechanism for virializing baryonic and warm dark matter into disks.  These cosmic strings survive as configurations which thread spiral and elliptical galaxies leading to the observed flatness of rotation curves and the Tully-Fisher relation.  We find a relationship between the rotational velocity of the galaxy and the string tension and discuss the testability of this model.}
\end{abstract}

\maketitle
\section{Introduction}

A number of independent observations have lead astronomers to the the conclusion that roughly one third of the 
universe is dominated by a cold, collisionless, non-baryonic dark matter.  Nothing is known about dark matter aside from the 
observed properties we have attributed to it such as flat rotation curves of galaxies, lensing and Large Scale Structure (LSS) \cite{krauss}.

Another aspect of the dark matter problem arises from the correct magnitude of the power-spectrum for LSS during the  recombination epoch.  The power spectrum is nearly scale invariant and a form of CDM is necessary to explain the magnitude of the gravitational potential necessary for LSS \cite{robert}.   In the contemporary universe, dark matter is needed to explain the observed flat rotation curves, which is thought to exist in the configuration of a halo.  Clearly, whatever dark matter is, it should account for both phenomena.  Usually it is thought that after matter-radiation equality, the density contrast of DM clusters and creates a potential well so that baryonic matter can fall into it.  There are a number of experiments that will try to directly detect DM.  Nonetheless, it is necessary to explore all possible explanations of the dark matter problem.  In this paper we shall consider a different scenario for dark matter which we feel is well motivated in the context of fundamental physics.

We propose a mechanism wherein cosmic strings play a central role in explaining dark matter both at large and galactic/cluster scales.  It is well known that cosmological symmetry breaking phase transitions can occur in the past and topological defects will be topologically guaranteed depending on the symmetry breaking pattern in the vacuum manifold.  

In this model we propose that a rigid network of cosmic strings form after the inflationary epoch \footnote{Generically cosmic strings may form after phase transitions in stringy alternatives to inflation, and this mechanism may also apply in these models \cite{st,henry,bv,abe}}.  Because they are very stiff, the cosmic strings have vanishing peculiar velocity with respect to the comoving frame.   These strings have wiggles and are endowed with an attractive radial force leading to an efficient accretion mechanism for both warm dark and baryonic matter.   Due to inflationary initial conditions, both the strings and the baryonic matter are spatially distributed in a scale invariant manner.  

The baryonic matter will accrete onto the core of the cosmic string forming a bound system.  The cosmic string may generate the necessary large scale power to account for dark matter by either the emission of dark matter, such as axions (if the strings are axionic strings) to provide the necessary power on cluster and supercluster scales.    At later times,  the system evolves into a nonlinear regime of the contemporary universe and the network of cosmic strings will come to dominate matter density of universe.  As a result one can relate the string tension to the observed flat rotation curve.

\section{Cosmic Strings and the Early Universe}
It was shown that as the Universe cools to a critical temperature $T_{c}$ associated with spontaneous symmetry breaking, various species of topological defects can form.  If the symmetry is broken from a group G to a subgroup H, the manifold of degenerate ground states is $\cal{M}\rm ={G \over H}$ and the topology of this manifold determines the types of topological defects that can form.  In particular if non-contractible loops exists in the vacuum manifold or equivalently, if there is a non-trivial first homotopy group $\pi_{1}(\cal{M}\rm) \neq 0 $, then stable cosmic strings can exist\cite{mark}.  

Cosmic strings were postulated in the 80's to be an alternative scenario for LSS, but were eventually ruled out by Boomerang and WMAP measurements of the peaks in the CMB angular power spectrum.  These peaks can be traced back to the fact that super horizon perturbation fourier modes are in phase at the end of inflation.  The positions and heights of these peaks are very characteristic of the inflationary scenario.  The CMB experiments do not, however, prove that cosmic strings do not exist in the universe.  It only proves that they do not contribute more than an insignificant proportion of the primordial density perturbation.  The CMB observations give an upper limit on the value of the parameter $G\mu$.  Pogosian et al \cite{levon} find an upper limit 
\be
G\mu \leq 1.3\times 10^{-6}\sqrt{\frac{B\lambda}{0.1}}, 
\ee
where $\lambda$ is the probability of intercommuting when two strings meet and $B$ is the fraction of the CMB power spectrum attributable to cosmic strings which is around .1.    

There are two important loopholes that will be crucial for our scenario. First, in the context of inflation, strings can play a role in seeding structure\cite{joao}.  It has been shown that most inflationary models motivated from grand unified theories and string theory generically produce cosmic strings\cite{Hindmarsh:1994re,Copeland:2003bj}.  Therefore cosmic strings can work together with inflation as a means of generating the power on large scales necessary for sufficient structure formation.
\section{The Mechanism}

At the end of inflation cosmic strings will form according to \cite{Jeannerot:2003qv}.  Inflation plays an important role in the dark matter problem since the power necessary for LSS at horizon crossing is best understood from the density fluctuations generated during inflation, rather than cosmic strings alone. Nonetheless, it is plausible that these fluctuations are sourced by cosmic strings which formed during inflation, or by a curvaton field associated with cosmic strings \cite{Endo:2003fr}.   But once the string network has formed we will show that on the scales of galaxies, there is no need for dark matter to explain the flat rotation curves.

The most important part of the model is the realization that if strings pierce the center of galactic planes, flat rotation curves with the correct magnitude arise without fine tuning.   When strings form in the early universe, there is a net radially directed force on the baryonic matter toward the core of the string.  The baryonic matter eventually undergoes rotation about the string core due to virialization.  This condition relies on the domination of the attractive potential of the string core over the other forces acting on the baryons, including self gravity and the background expansion.  We will provide some estimates of infall mechanism in the next section.  

Infinitely straight strings are known to have a Minkowski metric with a deficit angle, therefore they admit straight geodesics far away from the string.  However, this is not the case for strings with wiggles as they have an attractive potential in the radial direction.   The energy momentum tensor of a wiggly global string segment oriented the z axis is approximated as:

\be \label{emtensor} T^{\mu}_{\nu} =\delta(x)\delta(y)diag(\mu,0,0,T) \ee
where $\mu$ and $T$ are the effective mass per unit length and tension of the string respectively.
In the case with $T=\mu$, the above is the effective energy momentum tensor of an unperturbed string with tension $\mu$ as seen from distances much larger than the thickness of the string (a Nambu-Goto string).  However, real strings develop small-scale structure and are not well described by the Nambu-Goto action.  When perturbations are taken into account $T$ and $\mu$ are no longer equal and can only be interpreted as effective quantities for an observer that cannot resolve the perturbations along its length.  In this case, Carter has proposed \cite{Carter:1994zs}that when the transverse speed, $c_{T} = (T/\mu)^{-1/2}$ and longitudinal speed $c_{L} = (-dT/d\mu)^{1/2}$ coincide, there is a new equation of state
\be \mu T=\mu_{0}^{2}, \ee
hence, describing the energy-momentum of a wiggly cosmic string as seen by an observer that cannot resolve the wiggles or other irregularities.

The gravitational field of the wiggly string can be found by solving the linearized Einstein equations with $T^{\mu}_{\nu}$  from eq (\ref{emtensor}).  This gives \cite{alex}
\be h_{00}=h_{33}=4G(\mu-T)ln(r/r_{0}) \ee
\be h_{11}=h_{22}=4G(\mu+T)ln(r/r_{0}) \ee
where $h_{\mu\nu}=g_{\mu\nu}-\eta_{\mu\nu}$ is the metric perturbation and $r_{0}$ is a constant of integration \cite{brandon}.
In this gauge we can obtain the gravitational force acting on a test particle follows
\be \vec{\nabla}\Phi= \frac{G(\mu-T)}{r} \ee
where $2\Phi=h_{00}$.

Now let us imagine placing in this string perpendicular to the center of the disk of a spiral galaxy\footnote{The metric of the cosmic string is determined in a cylindrical coordinate system, while for a galaxy, spherical polar coordinate system is used.  Therefore, if one assumes that a galaxy forms where two strings cross, then it is a good approximation to use the $log(r/r_{r})$ potential in the galactic coordinate system.}.  In the weak field limit the modified poisson equation is:
\be \label{mine} a_{tot}= \vec{\nabla}\Phi_{Galaxy} + \vec{\nabla}\Phi_{CS}=  \frac{GM}{r^{2}} + \frac{G(\mu - T)}{r} \ee

If one modifies Newton's law to the following form \cite{Aguirre:2003pg}:
\be \label{anth} a=\frac{GM}{r^{2}} + \frac{BM^{1/2}}{r} . \ee
where $B=\sqrt{G a_{0}}$ and $a_{0}=cH_{0}$ and $r$ is the radius in spherical polar coordinate such that the origin lies at the center of a galaxy, we will obtain both flat rotation curves as well as 
the observed Tully-Fisher (TF) for spiral galaxies (ie. $L \propto v^{4}$ where L is the luminosity).  
Let us compare eqs (\ref{mine}) and (\ref{anth}) we immediately see that they are identical if 
\be \label{sq} G(\mu - T) =\sqrt{Ga_{0}M} \ee
We can find a condition on the string mass and the galaxy mass in the regime where the string potential dominates, from eq \ref{mine}:
\be \mu - T=\frac{M_{g}}{R_{g}} \ee
where $R_{g}$ is assumed to be the radius of the Galaxy.
We now substitute this value into eq(\ref{sq}) and find the relation
\be \frac{GM_{g}}{R_{g}^{2}}=cH_{0} \ee
which is the phenomenological relation of Milgrom's law (MOND) \cite{mond}.  We have therefore found a consistency between our cosmic string configuration and the phenomenological
fit for spiral galaxies which lead to flat rotation curves and TF.   

Our result states that the tension of the string possesses the correct magnitude to account for the flatness of the rotation curve.   It is intriguing to ponder the nature of the dynamical origin of this relationship; it is expected to be nonlinear in origin.  Furthermore the magnitudes match because WMAP constrains $G(\mu- T) \leq 10^{-7}$ and it is observed that $\frac{GM_{g}}{R_{g}} \sim 10^{-7}$\cite{frenk}.
 
\subsection{Cosmic Strings and Structure Formation Revisited}
The CMB observation that $\Omega=1$ on large scales necessitates a dark component which clusters on large scales.  However, at the same time it has been reported that the WMAP data requires loss of power at large angular scales.  Cosmic string networks can possibly account for both phenomena.  First, it was shown by Vilenkin that an isotropic network of nonintercommuting strings can solve the $\Omega=1$ problem of the inflationary scenario\cite{alex}.  This mainly happens because the network energy density $\rho_{n} \propto a(t)^{-2}$, leads to the expansion rate $a(t) \sim t$.  The expansion rate is the same as a matter-filled universe with $\Omega <1$ at late redshifts $2 +z <<\Omega^{-1}$.  Thus the evolution of a string-dominated universe with $\Omega=1$ mimics the behavior of an $\Omega \sim (2+z_{s})^{-1} <1 $ universe\cite{Vilenkin:1984rt,Spergel:1996ai} \footnote{An $\Omega <1$ universe has problems explaining the observed location of the first acoustic peak in the CMB data, we shall discuss this issue in the context of rigid stings in future work.} .   This could potentially solve the $\Omega=1$ dark matter problem as well as providing a beneficial increase in the apparent age of the universe.   For the strings not to spoil nucleosynthesis, there will be a constraint on the correlation length of strings $\xi$.   This constraint originates from demanding the strings do not overclose the universe:
\be \Omega_{s}=\frac{\rho_{s}}{\rho_{c}} < 1\ee
Using the solution for cosmic strings we arrive at the following result.
\be \large[\frac{\xi}{t}\large]_{now}=(30G\mu)^{1/2} \ee
For our model $G\mu \sim 10^{-7}$ leads to a correlation length \be \xi \sim 1 Mpc \ee
 
Notice that if the strings are formed at the end of inflation 
and the network of strings are frozen in comoving coordinates from the beginning, then the 
correlation length will be much smaller than $1 Mpc$; which is comparable to the comoving Hubble scale 
during recombination. This happens primarily because the cross section of typical field theoretic string interactions is taken to be a constant.  However, in the context of 
cosmic superstrings the cross section can decrease with time, since the dilaton would not be constant\cite{joe,anupam}.  If the string network evolves from the end of inflation to recombination
and freezes, then the correlation length can be on the order of $1 Mpc$ today. 
 
This model was considered in the past by Vilenkin, but it failed to explain clustering on galaxy-cluster scales.  Therefore, provided that this network can accrete baryonic matter on smaller scales, both large and small scale dark matter problems can be solved.   Below we present the small scale solution in the spirit of past work by Zanchin, Lima and Brandenberger (ZLB) \cite{Zanchin:1996cp}.
\subsection{The ZLB Mechanism and Rigid Strings}
If indeed cosmic strings are at the center of galaxies we must understand how galaxies formed in that manner; how matter fell into the core of the strings at high redshifts.   In past investigations cosmic strings were proposed as the seeds for galaxy formation.  There were three distinct mechanisms: wakes, cusps and filaments.   We shall focus our attention on the filamentary mechanism.  The basic idea is that slowly moving wiggly cosmic strings have a locally attractive force which can accrete matter in its wake.  This happens because strings with small scale structure (wiggles) have a local Newtonian potential; discussed in Section III.  Therefore, our mechanism will require matter to accrete onto the string core due to the attractive potential; the same one which yields the flat rotation curve today.  If the cosmic strings are rapidly fluctuating then the matter is less likely to fall in. So what we need is a mechanism which suppresses relativistic transverse fluctuations of the string locally.   

An attractive way to maximize infall of hot dark matter and baryonic matter and suppress relativistic transverse oscillations is for strings to behave like a solid.   Originally cosmic strings in structure formation had an equation of state only relating pressure and energy density.  But it was shown by Bucher et al. that strings can be have a bulk modulus which renders them stiff\cite{Bucher}.   In this case there will be shear stress which acts to prevent the string network from undergoing large transverse oscillations.  Bucher and Spergel realized that a static isotropic distribution of strings could have a large bulk rigidity modulus \cite{Bucher}.  The rigidity bulk modulus is defined as the change in energy density $\rho$ due to an infinitesimal volume conserving spatial displacement
\be \delta \rho = \mu e^{ij}e_{ij} \ee
where $e_{ij}$ is an infinitesimal strain tensor and $\mu$ is the bulk rigidity modulus.  In a recent paper, Bucher, Carter and Spergel found that
networks of strings and domain walls can have a large bulk modulus
\be \mu =\frac{4}{14}\rho \ee
This large bulk modulus for a network of solid cosmic strings and domain walls can suppress transverse fluctuations\cite{carter,Bucher}.  With this in mind we are equipped 
to resort to the accretion mechanism proposed by Brandenberger et al. \cite{Zanchin:1996cp}.  The mechanism demonstrates that strings carry a substantial amount of small-scale structure, thereby acting gravitationally as a Newtonian line source whose effects dominate the velocity perturbations.  Even with hot dark matter, the first nonlinear filamentary structures form at a redshift close to $100$, and there is sufficient nonlinear mass to explain the observed abundance of high redshift quasars and damped Lyman alpha systems.   The mechanism makes use of the Zel'dovich approximation.  

Let us follow a heuristic argument put forth by Shellard and Vilenkin.  Consider a hot dark matter particle whose position is given by
\be r(q,t) = a(t)[q-\psi(q,t)], \ee
where $q$ is the comoving coordinate and $\psi$ is the comoving displacement due to the gravitational perturbation.   By combining Newtonian gravitational force equation, the Poisson equation for the Newtonian gravitational potential, using conservation of matter and expanding to first order in $\psi$ one arrives at the following equation for the comoving displacement of a background particle
\be \frac {\partial^{2}\psi}{\partial t^{2}} + \frac{2\dot{a}\partial\psi}{ a\partial t} + \frac{3\ddot{a}}{a}\psi= S(q,q'), \ee
where $S$ stands for the source term.  So far the analysis is general.  We now specialize to our case in which the perturbations are generated by a long straight string with small scale structure whose strength is given by the value of $G\lambda$, where $\lambda= \mu-T$.  Then the source term is given by
\be S=\frac{2\lambda G}{a^{2}}\frac{q-q'}{|q-q'|^{2} }. \ee
In the case of hot dark matter, the authors  \cite{Brandenberger:1987er, Zanchin:1996cp} showed that 
at a "turn around" time, the particles decouple from the Hubble flow and start falling back towards the string core.  The ZLB mechanism predicts that non-linear structures form at a redshift close to $z \sim 100$ with sufficient nonlinear mass to explain the observed abundance of high redshift quasars and damped Lyman alpha systems.   

\section{Discussion}

In this paper I speculate that flat rotation curves can be accounted for if cosmic strings thread the center of galaxies.  These strings would form a web like structure on large scales.  I also argued that these string networks could form generically after a $U(1)$ symmetry breaking after inflation.   The cosmic string would accrete baryonic matter into its core as a mechanism for virializing protogalaxies.  I revisited this accretion mechanism of Brandenberger and Turok and the ZLB mechanism, using the Zeldovich approximation.  On large scales I calculated the persistence length of the cosmic string network given the constraints on the tension of the string, which was found be consistent with large scale structure data.  

This idea is new and will have to pass two tests.  First, this network is non-relativistic so the current weak lensing constraints on cosmic strings will not apply.  Nonetheless, a weak lensing analysis needs to be carried out for such a network of strings.  Second, a concrete microphysical mechanism is lacking to explain the needed bulk rigidity of the sting network.  These two issues are currently in pursuance.

This model is testable since it predicts no excess potential in the z-direction
 for thin disk spirals; the z-component of the velocity
 dispersion should trace the baryonic mass, not the
 dark mass.  The dominant mode for forming ellipticals is through spiral collisions - what do the strings do then?  This situation could be analyzed computationally.
 In general the strings will be misaligned, which is good in terms of
 inducing high velocity dispersions \footnote{I thank Steinn Sigurdson for pointing this out.}.

 How does this model impact Dwarf galaxies? Do they also have strings? Dwarf galaxies
 have an unexplained scale length of $~ 300pc$
 and there are far fewer than expected from standard CDM models, but
 there are still many of them for each big galaxy and
 they fall into big galaxies, especially at $z \gg 2$.

Since most galaxies possess a supermassive black hole(SMBH), are these strings tied
 to the SMBH's rotation pole?
If so, then this model predicts that central SMBHs must have their spins aligned with the galaxy.
 
\section{Acknowledgements}
I give special thanks to Michael Peskin for his feedback and encouragement in the early stages of this project.  I also wish to thank Niayesh Afshordi, Robert Brandenberger, Edward Baltz, Alex Vilenkin, Levon Pogosian, Steinn Sigurdson and Robert Wagoner for insightful discussions.

 \vskip 0.5 true in

\end{document}